\documentclass[twocolumn,showpacs,aps,amsmath,amssymb]{revtex4}
\usepackage{psfrag,graphicx}
\usepackage[labelsep=space,justification=raggedright,font=normalsize]{caption}

\newcommand{\hrm}{\mathrm{h}}
\newcommand{\thrm}{\mathrm{th}}
\newcommand{\Nrm}{\mathrm{N}}

\begin{document} 

%Title of paper
\title{Shape of primary proton spectrum in multi-TeV region\\ from data on vertical muon flux}
\author{A.~V.~Yushkov\footnote{yushkov\_av@mail.ru}}
\affiliation{Sezione INFN di Napoli, I-80126, Napoli, via Cintia, Italy}
\author{A.~A.~Lagutin}
\affiliation{Altai State University, 656049, Barnaul, Lenin pr., Russia}

\date{\today}

\begin{abstract}
It is shown, that primary proton spectrum, reconstructed from sea-level and
underground data on muon spectrum with the use of QGSJET~01, QGSJET~II,
NEXUS~3.97 and SIBYLL~2.1 interaction models, demonstrates not only
model-dependent intensity, but also model-dependent form. For correct
reproduction of muon spectrum shape primary proton flux should have non-constant
power index for all considered models, except SIBYLL~2.1, with break at energies
around 10--15~TeV and value of exponent before break close to that obtained in
ATIC-2 experiment. To validate presence of this break understanding of inclusive
spectra behavior in fragmentation region in p-air collisions should be improved,
but we show, that it is impossible to do on the basis of the existing
experimental data on primary nuclei, atmospheric muon and hadron fluxes.
\end{abstract}

\pacs{95.85.Ry, 96.50.sb}

\maketitle

\section{Introduction}
At present information on the characteristics of hadronic interactions in
fragmentation region is still scarce or missing and experiments with `roman pots'
are anticipated to improve the situation. Some of this information, in principle,
could be obtained with the use of the data on cosmic ray (CR) muon and hadron
spectra, provided primary spectra are known with high precision, but that is not
the case. The obvious obstacle here is that at high energies primary cosmic ray
(PCR) fluxes, measured in direct experiments, themselves are functionals of
various interaction parameters plus their accuracy is appreciably affected by
additional systematic
effects~\cite{grigorov1973_eng,jacee1986,msuhadrons2000_eng,runjob_eng}. In the
series of our papers~\cite{ya2004,ya_yaf2006_deficit_eng} we underlined, that
these effects can lead to underestimation of light nuclei fluxes and thus can
explain discrepancy between measured and calculated muon fluxes for
$E_\mu>100$~GeV. Preliminary data of ATIC-2~\cite{atic2_RCRC2007}, covering the
gap between magnetic spectrometer and  emulsion chamber experiments, seems to be
in concordance with our conclusions, but situation is more complicated in fact,
as further consideration will show. ATIC-2 slope of proton spectrum
$\gamma_p=2.63$~\cite{zatsepin_3cmodel} for primary energies below 10~TeV is in
remarkable contradiction with previously measured values $\gamma_p=2.74$ and
$\gamma_p=2.80$ by RUNJOB~\cite{runjob_apj2005} and JACEE~\cite{jacee}
experiments correspondingly, but this discrepancy is removed by steepening in
ATIC-2 proton spectrum at energies above 10~TeV. These new data were already
exploited in extensive calculations of muon and hadron fluxes with number of
interaction models at different atmospheric depths and zenith angles
in~\cite{kochanov_2008} where it was shown, that their use allows to get
reasonable agreement with the most of the data under appropriate choice of
hadronic interaction parameters. In fact, it is not possible to obtain concordant
conclusions on primary spectra or hadronic interactions parameters coming even
from much smaller subset of the experimental data. In this paper we demonstrate
this on the basis of analysis of the data on muon flux at vertical direction and
only one set of the data on hadron flux of EAS-TOP~\cite{eas_top_p}. First, we
use the data of sea-level and underground experiments to obtain conclusions on
behavior of muon spectrum at sea-level in the energy range 40~GeV--10~TeV.
Further we analyze influence of uncertainties in the muon data and interaction
parameters on properties of reconstructed primary proton fluxes. And finally we
show, that such different in approach and characteristics interaction models as
SIBYLL~2.1~\cite{sibyll,sibyll2.1} and NEXUS~3.97~\cite{nexus} can bring to
hardly distinguishable predictions on muon and hadron fluxes.

\section{Sea-level muon spectrum from underground experiments data}

\begin{figure}[bt]
\centering\includegraphics[width=.48\textwidth]{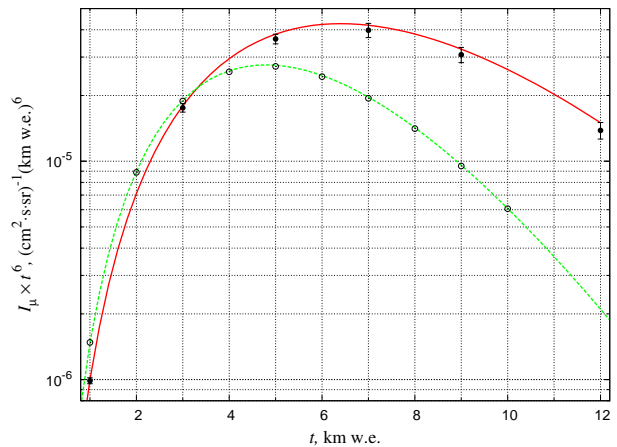}
\caption{(color online). Muon intensity in water and in standard rock. Water
(muon spectrum at sea level from~\cite{mum}): solid line --- present work; closed
circles~---~MUM~\cite{mum}, error bars show uncertainty due to $\pm1$\% variation
of total muon energy losses. Standard rock (muon spectrum at sea level
from~\cite{lvd99}): dashed line --- present work, open
circles~---~MUSIC~\cite{kudryav_mu_n}.} \label{fig:vsmum}
\end{figure}

Depth-intensity relation, needed for reconstruction of sea-level muon spectrum,
may be obtained via numerical solution of one-dimensional transport equation. In
adjoint approach this equation has the following form~\cite{lagutin_prepr94}
\begin{eqnarray*}
\frac{\partial \bar q(t,E)}{\partial t}+\sigma\bar q(t,E)
-\sum_\beta\int\limits^E_{E_\mathrm{th}} dE'W_\beta(E,E')\bar q(t,E')=\\=D(t,E).
\end{eqnarray*}
Here $\bar q(t,E)$~---~is survival probability of muon with energy $E$, being
born at the distance $t$ from detector, $\sigma$~---~total interaction
cross-section, $W_\beta(E,E'),\ \beta=i,r,p,h$~---~differential cross-sections
for processes of ionization, bremsstrahlung, pair production and photonuclear
interaction correspondingly, $D(t,E)$~---~detector sensitivity function. The
numerical method, applied for solution of this
equation~\cite{ya_yaf2006_zemlya_eng}, allows to avoid any approximations (such
as continuous losses one) and to obtain muon intensities at large depths of
matter with account of fluctuations in all muon interaction processes. Accuracy
of our calculations of muon survival probabilities and intensities was thoroughly
examined in~\cite{ya_yaf2006_zemlya_eng} and comparison with the results of
Monte-Carlo codes MUM~\cite{mum} and MUSIC~\cite{music} is presented in
Fig.~\ref{fig:vsmum}. Anticipating further discussion of sea-level muon spectrum
behavior it is necessary to note, that our calculations give upper estimate of
muon flux at large depths in comparison with MUM, because of use of $\sim$1\%
lower muon energy losses~\cite{ya_yaf2006_zemlya_eng}.

\begin{figure*}
\centering\includegraphics[width=0.49\textwidth]{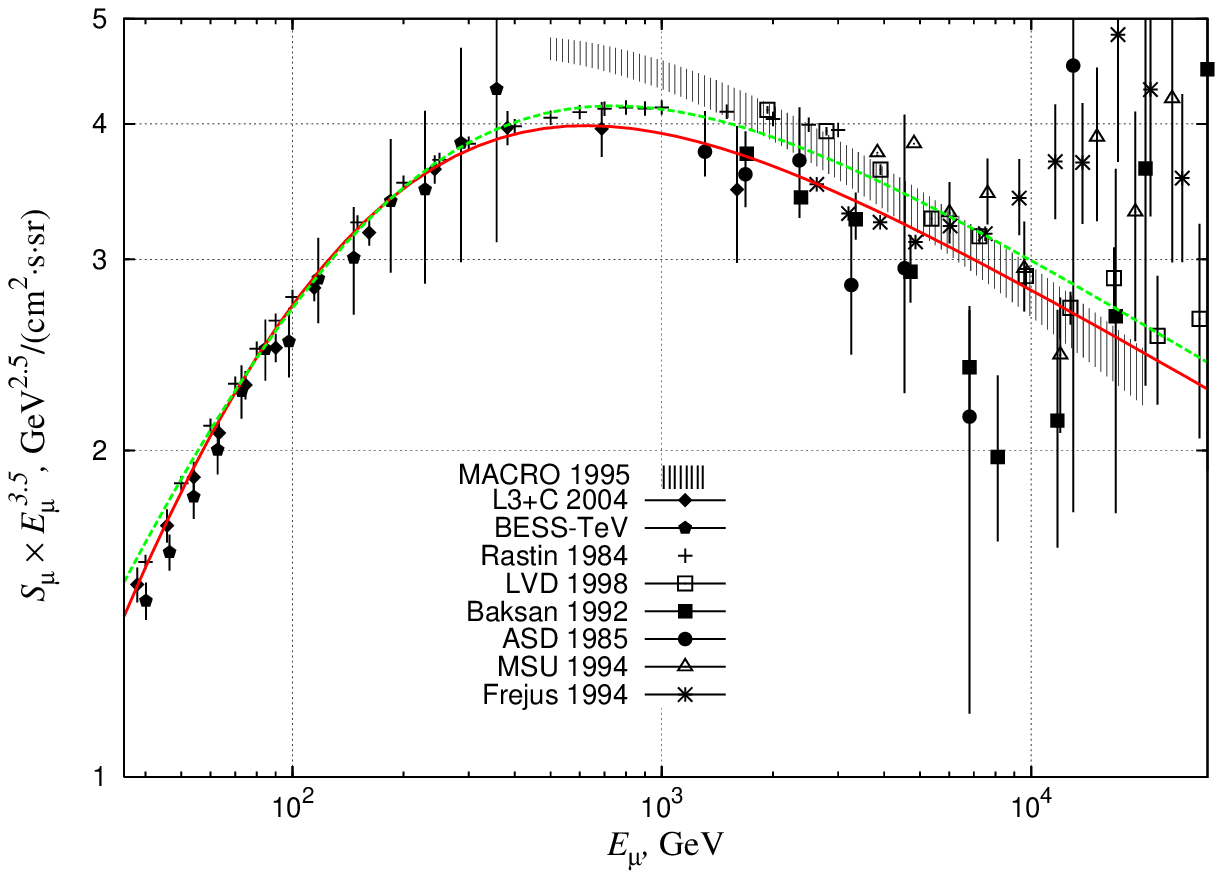}
\centering\includegraphics[width=0.49\textwidth]{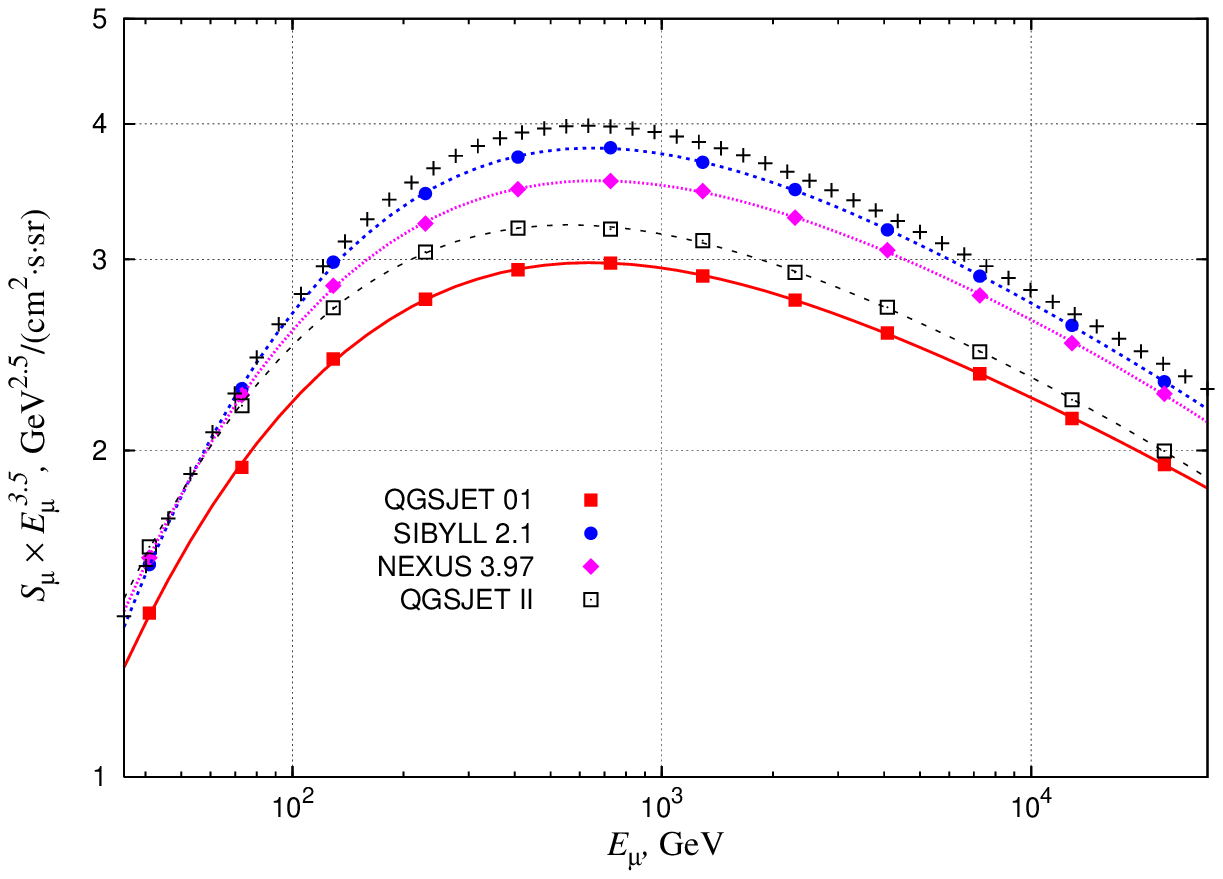}
\caption{(color online). Sea-level muon spectrum. Experimental data:
\cite{macro}~MACRO~1995, \cite{l3_c2004}~L3+C~2004, \cite{bess_tev}~BESS-TeV,
\cite{rastin}~Rastin, \cite{lvd}~LVD~1998, \cite{baksan92_eng}~Baksan~1992,
\cite{asd_eng}~ASD~1985, \cite{msu_mu_eng}~MSU~1994, \cite{frejus}~Frejus~1994.
{\em Left:} Solid line~---~muon spectrum~(\ref{eq:bogdanova}), dashed
line~---~muon spectrum~(\ref{eq:mbogdanova}). {\em Right:} Muon spectra at sea
level for PCR spectra from~\cite{gaisser2002} with high helium flux. Muon
spectrum parametrization~(\ref{eq:bogdanova}) is shown with crosses.}
\label{fig:param}
\end{figure*}

\begin{figure}
\centering\includegraphics[width=.48\textwidth]{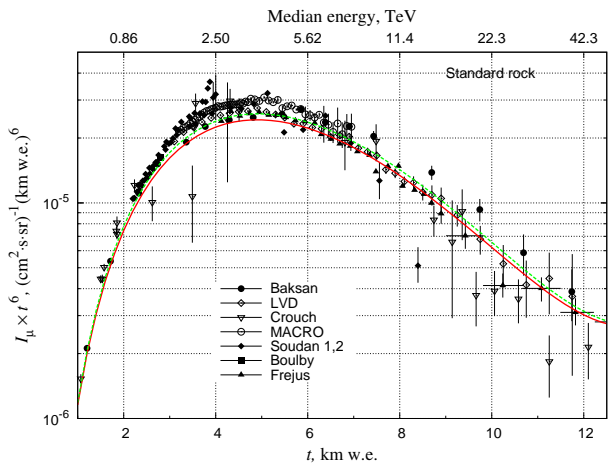}
\caption{(color online). Vertical muon intensity in standard rock. Experimental
data: \cite{baksan87}~Baksan, \cite{lvd}~LVD, \cite{crouch87}~Crouch,
\cite{macro}~MACRO, \cite{soudan1,soudan2}~Soudan 1,\,2, \cite{boulby}~Boulby,
\cite{frejus,frejus90}~Frejus. Present work calculations: solid line~---~for muon 
spectrum~(\ref{eq:bogdanova}), dashed line~---~for muon
spectrum~(\ref{eq:mbogdanova}). Neutrino induced muon contribution is taken
from~\cite{lvd_mu_nu}.}\label{fig:strock}
\end{figure}

To describe the data on muon intensity underground and at sea-level, and to
estimate influence of uncertainty in muon flux data on reconstruction of primary
proton flux, we used two parameterizations in the simple form, proposed in work
of Bogdanova~et~al.~\cite{bogdanova2006}. Original fit for the vertical
from~\cite{bogdanova2006}
\begin{equation}
\label{eq:bogdanova}
S_\mu(E)=18/(E+145)/(E+2.7)^{2.7},\ (\text{cm}^2\cdot\text{s}\cdot\text{sr}\cdot\text{GeV})^{-1},
\end{equation}
provides good agreement with the data at sea-level (Fig.~\ref{fig:param}), but
leads to underestimation of the muon flux for the depths below 6~km~w.e.
(Fig.~\ref{fig:strock}).

To match better underground data for the depths 2--6~km~w.e. we shall also apply
modified fit with slightly $<10$\% increased intensity in multi-TeV region
\begin{equation}
\label{eq:mbogdanova}
S_\mu(E)=20.8/(E+194.3)/E^{2.71},\ (\text{cm}^2\cdot\text{s}\cdot\text{sr}\cdot\text{GeV})^{-1}.
\end{equation}
As it is seen from Fig.~\ref{fig:strock}, for depths from 4 km~w.e. up to 8
km~w.e., corresponding to $\sim2.5-10$~TeV median muon energies at sea level, 
use of this spectrum provides good agreement with the data of LVD~\cite{lvd},
BNO~\cite{baksan87} and Frejus~\cite{frejus,frejus90} collaborations and leads to
underestimation of data of MACRO~\cite{macro} and Soudan~\cite{soudan1,soudan2}
experiments.

Our consideration will touch muon energies only above 40~GeV, to exclude
different effects, not related to high-energy hadronic interactions features,
such as geomagnetic effect, influence of uncertainties in low-energy interaction
models, or even absorption of low energy muons in ground as in case of L3+C
detector.

\section{Calculations technique of atmospheric muon and hadron fluxes}
Average numbers of hadrons $N_\hrm(E_\Nrm,>E_\thrm)$ and muons
$N_\mu(E_\Nrm,>E_\thrm)$ with energies above $E_\thrm$ in EAS from primary
nucleon of energy $E_\Nrm$ were obtained with the help of one-dimensional
hybrid code CONEX~\cite{conex_pylos,conex_2006} in regime of cascade equations
solution for interaction models QGSJET~01~\cite{qgsjet},
SIBYLL~2.1~\cite{sibyll2.1}, NEXUS~3.97~\cite{nexus} and
QGSJET-II-03~\cite{qgsjetii,qgsjetiia,qgsjetiib}. To get differential spectrum
of hadrons (or in the same way of muons) for some energy $E_0$ the following simple formula had
been used:
\begin{equation}
\label{eq:diff}
S_\hrm(E_0)=\left[I_\hrm(>E_0-\Delta E)-I_\hrm(>E_0+\Delta
E)\right]/2\Delta E.
\end{equation}
Here $I_\hrm(>E_0)$ is the integral spectrum of hadrons
\[
I_\hrm(>E_0)=\int\limits_{E_0}^{E_\infty}S_\Nrm(E_\Nrm)N_\hrm(E_\Nrm,>E_0)dE_\Nrm
\]
for primary nucleon spectrum $S_\Nrm(E_\Nrm)$.

The interval width $\Delta E$ must be chosen to provide the difference between
integral intensities in~(\ref{eq:diff}) to be much larger then calculations
error. Test computations for $\Delta E=0.01E_0,0.02E_0,0.05E_0$ and observation
levels 820 and 1030~g/cm$^2$ for energy $E_0=1$~TeV brought to differential flux
values lying within $3\%$ from each other. Change of upper integration limit
$E_\infty$ in formula for $I_\hrm(>E_0)$ from $10^4\times E_0$ to $10^5\times
E_0$ gives less then 1\% increase of differential flux. And, the last, increase
of number of primary energy bins in $N_\hrm(E_\Nrm,>E_0)$ from 10 to 20 per order
introduces $\sim2\%$ variation to differential flux value. The listed error
sources partially compensate each other and the total error of our calculations
does not exceed 3\%. The calculation were performed for the set of energies,
coinciding with the set from EAS-TOP experiment paper~\cite{eas_top_p}.

\section{Sea-level muon fluxes}
As a basic model of PCR nuclei spectra the parameterizations
from~\cite{gaisser2002} were chosen. Nuclei with $A\geq4$ were treated in the
framework of the superposition model, high accuracy of this approach is well
known and was checked by our calculations with the use of CONEX both for muons
and hadrons once again.

\begin{figure*}
\centering\includegraphics[width=0.49\textwidth]{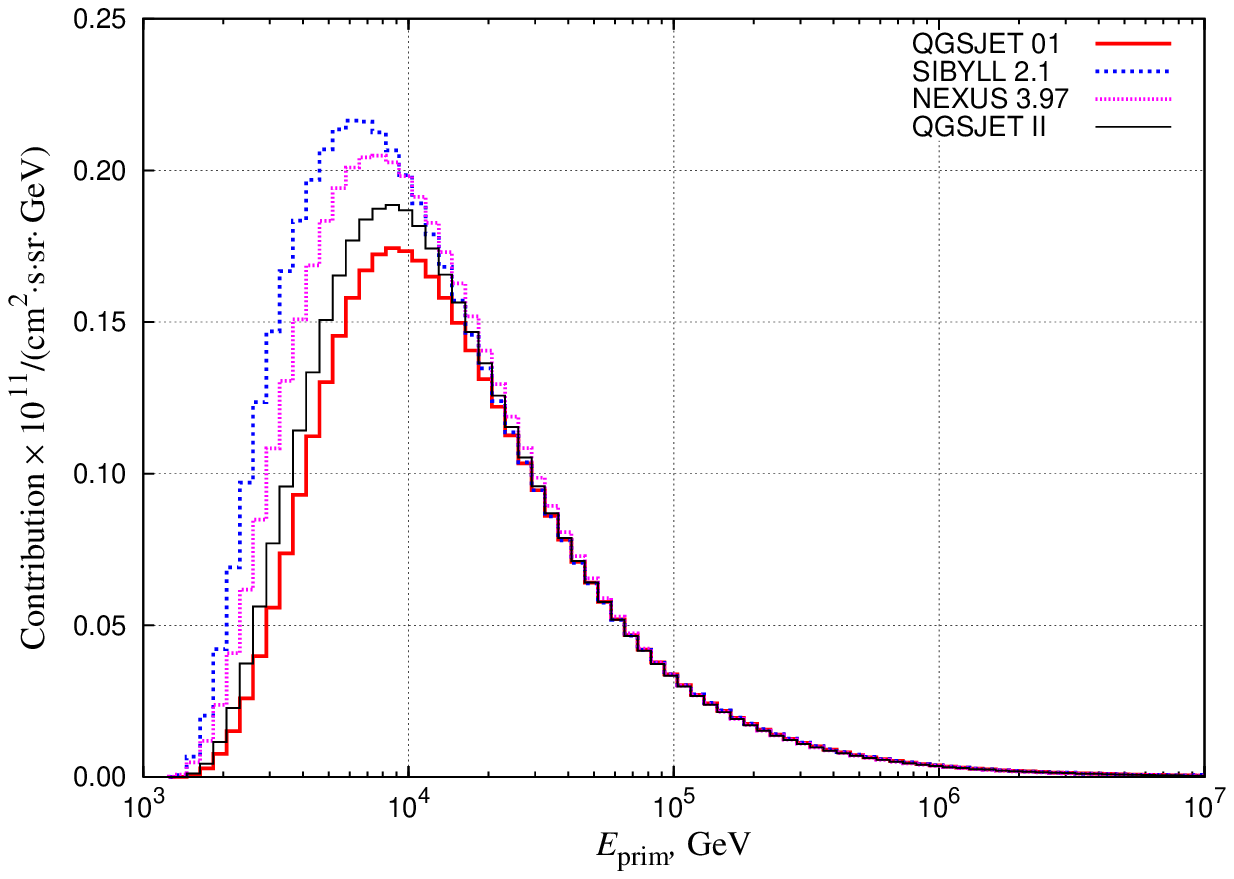}
\centering\includegraphics[width=0.49\textwidth]{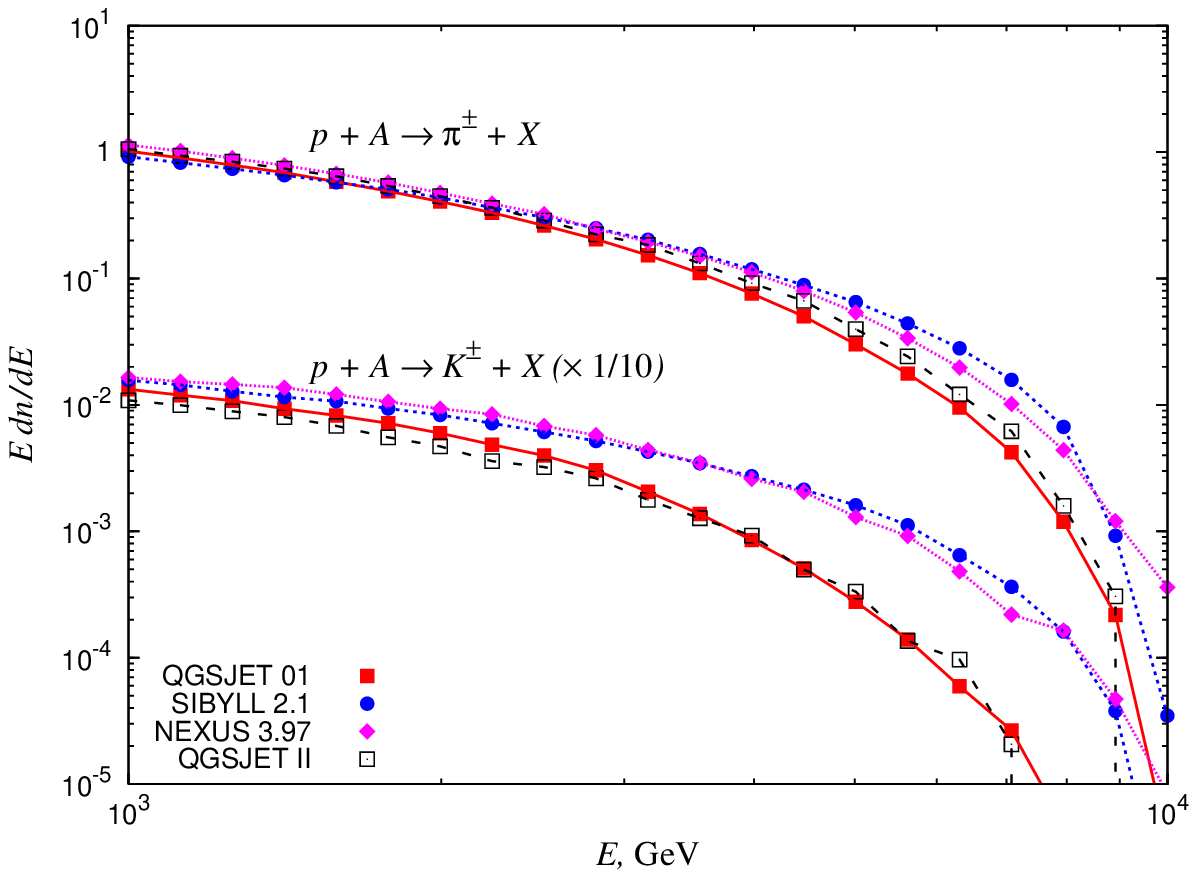}
\caption{(color online). {\em Left:} contribution of primary protons with
energies $E_\text{prim}$ to the muon differential spectrum at sea level for
$E_\mu=1.29$ TeV. {\em Right:} inclusive spectra $p+A\to \pi^\pm+X$ and $p+A\to
K^\pm+X$ (scaled down by 10) for incident proton with energy 10~TeV.}
\label{fig:inclusmu}
\end{figure*}

Comparison of the calculated muon fluxes with the experimental data, presented in
Fig.~\ref{fig:param}, reveal familiar picture of high energy muon deficit. The
reasons of its appearance were considered in our previous
papers~\cite{ya2004,ya_yaf2006_deficit_eng} and they still hold true
regardless of the fact, that three more interaction models were included in our
analysis. All interaction codes, except QGSJET~01, satisfactory describe data on
muon flux only up to $E_\mu\sim$100~GeV and then one by one fail to do it.
Accounting that such muon energies correspond to primary energies above 1~TeV,
studied with balloon(satellite)-borne emulsion chambers, we related muon deficit
to underestimation of primary light nuclei fluxes, taking place in these
experiments~\cite{ya2004,ya_yaf2006_deficit_eng}. Unfortunately,
disagreement between the models in the muon fluxes also appears at energies
around 100~GeV, thus making impossible precise reconstruction of primary nucleon
spectrum for $E_\text{prim}>1$~TeV. In fact, in such conditions there are no
reasons to rule out any of the models, except QGSJET~01, which leads to
remarkable disagreement with the experiment even in the range of reliable
magnetic spectrometers data on PCR and muon spectra.

To find out, why the models differ in the predicted muon fluxes let us consider
quite characteristic energy 1.29~TeV, where discrepancies between the models
reach appreciable values and the data on muons from underground installations are
yet quite reliable. Contributions of primary protons to the differential flux of
muons of the given energy, presented in Fig.~\ref{fig:inclusmu} show, that spread
in muon fluxes between the interaction models is entirely due to uncertainties in
the description of $\pi^\pm,K^\pm$-spectra in fragmentation region
$x=E_{\pi,K}/E_\text{prim}>0.1$. Since inclusive muon flux is sensitive nearly
only to the characteristics of the very first primary particle interaction,
hence, the harder these spectra are in the particular model, the larger muon
intensity its use leads to. For the lower values of $x$, i.e. for
$E_\text{prim}>10$~TeV, all the models give practically the same muon yields. As
noted above, in view of uncertain situation with primary spectra for
$E_\text{prim}>1$~TeV, one can not give preference to any of the models in
comparison with the others. If to demand the minimal disagreement with the direct
measurements data on PCR spectra, then obviously SIBYLL~2.1 satisfies this
requirement the best, or, on the other hand, one could say that it provides the
most acceptable description of $\pi^\pm,K^\pm$ production spectra in $p$-air
collisions in fragmentation region. We shall discuss this affirmation in detail
below.

\section{Proton spectrum from muon data}
From the previous consideration it is clear, that reconstructed fluxes of protons
shall be higher than measured in emulsion chamber experiments, but can be
comparable with recently obtained data of ATIC-2 group~\cite{atic2_RCRC2007}. We
performed the reconstruction simply by picking up of appropriate primary proton
flux parameters in order to minimize deviation of the obtained muon fluxes from
the parameterizations~(\ref{eq:bogdanova}) and (\ref{eq:mbogdanova}).

\begin{figure*}
\includegraphics[width=0.49\textwidth]{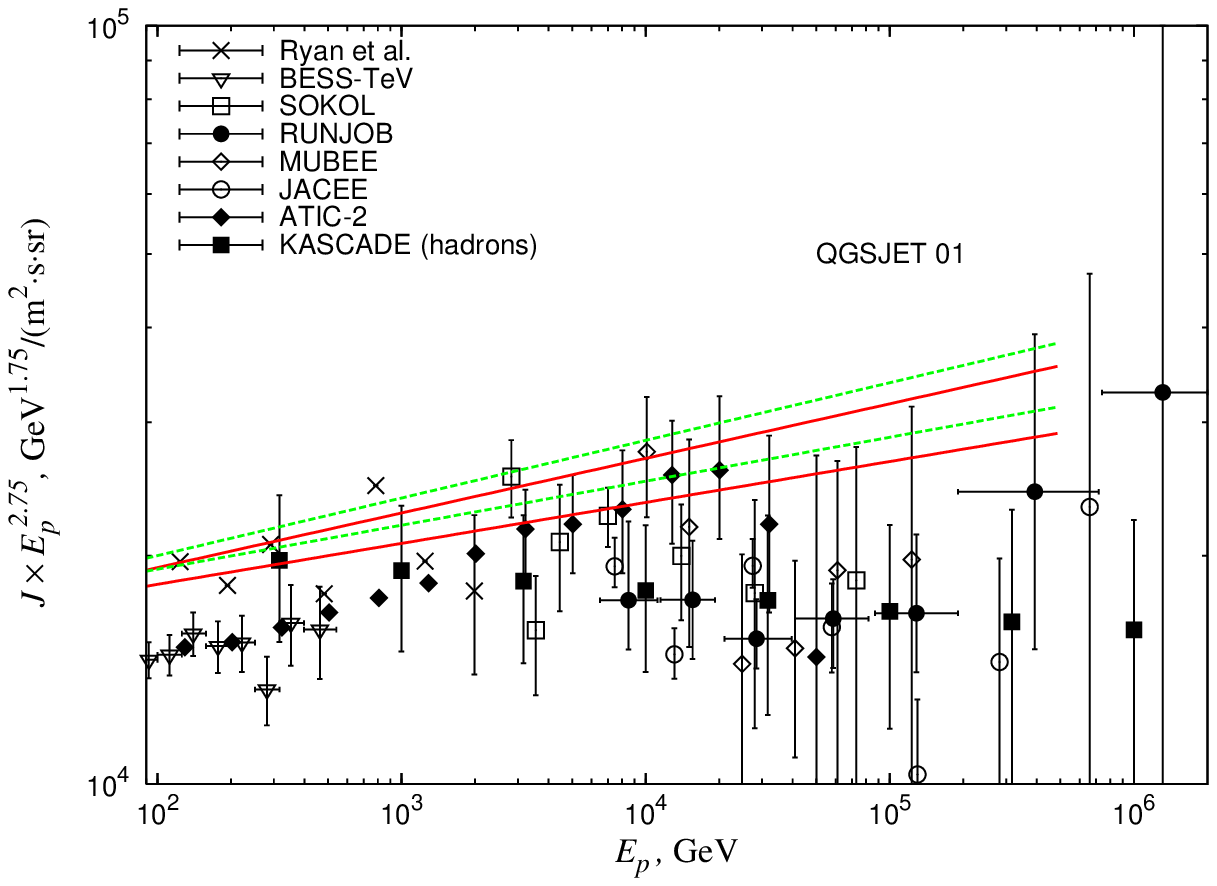}
\includegraphics[width=0.49\textwidth]{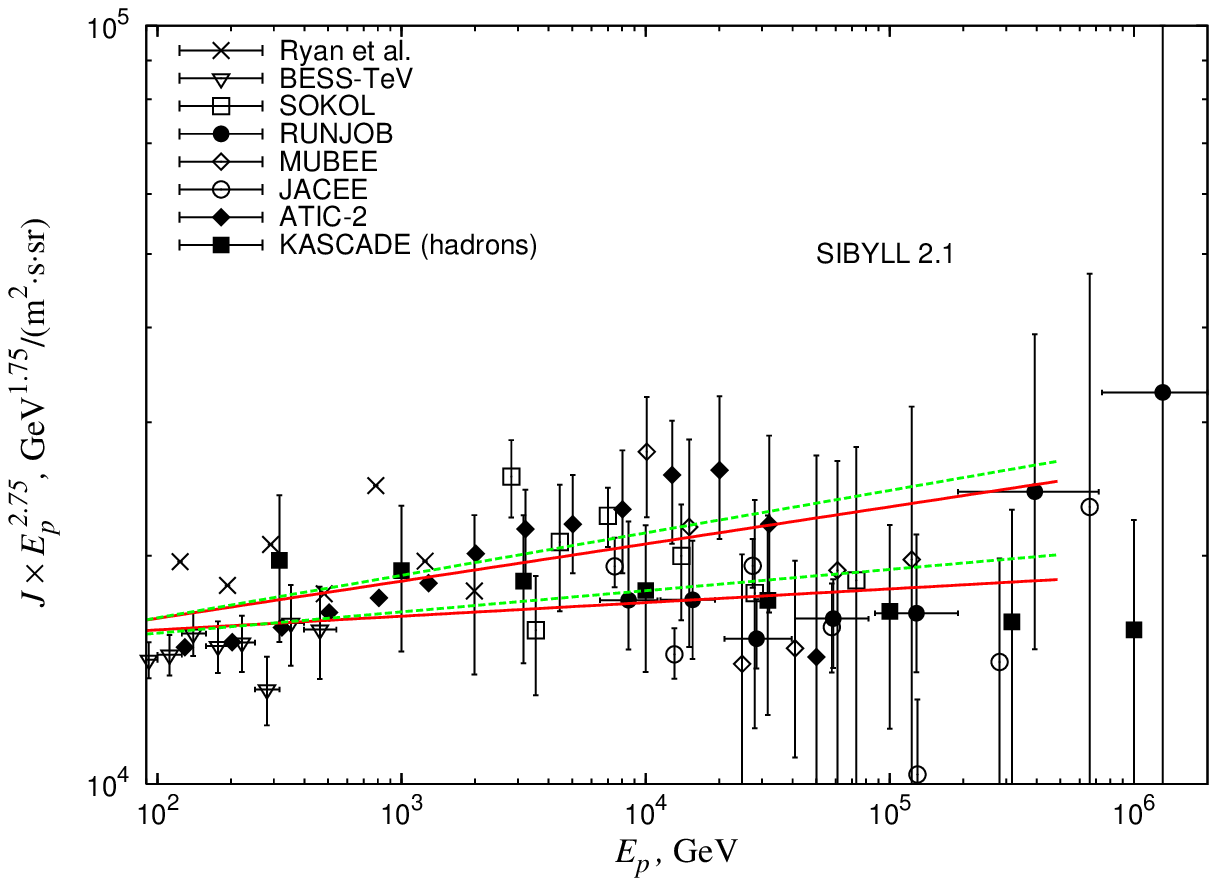}
\includegraphics[width=0.49\textwidth]{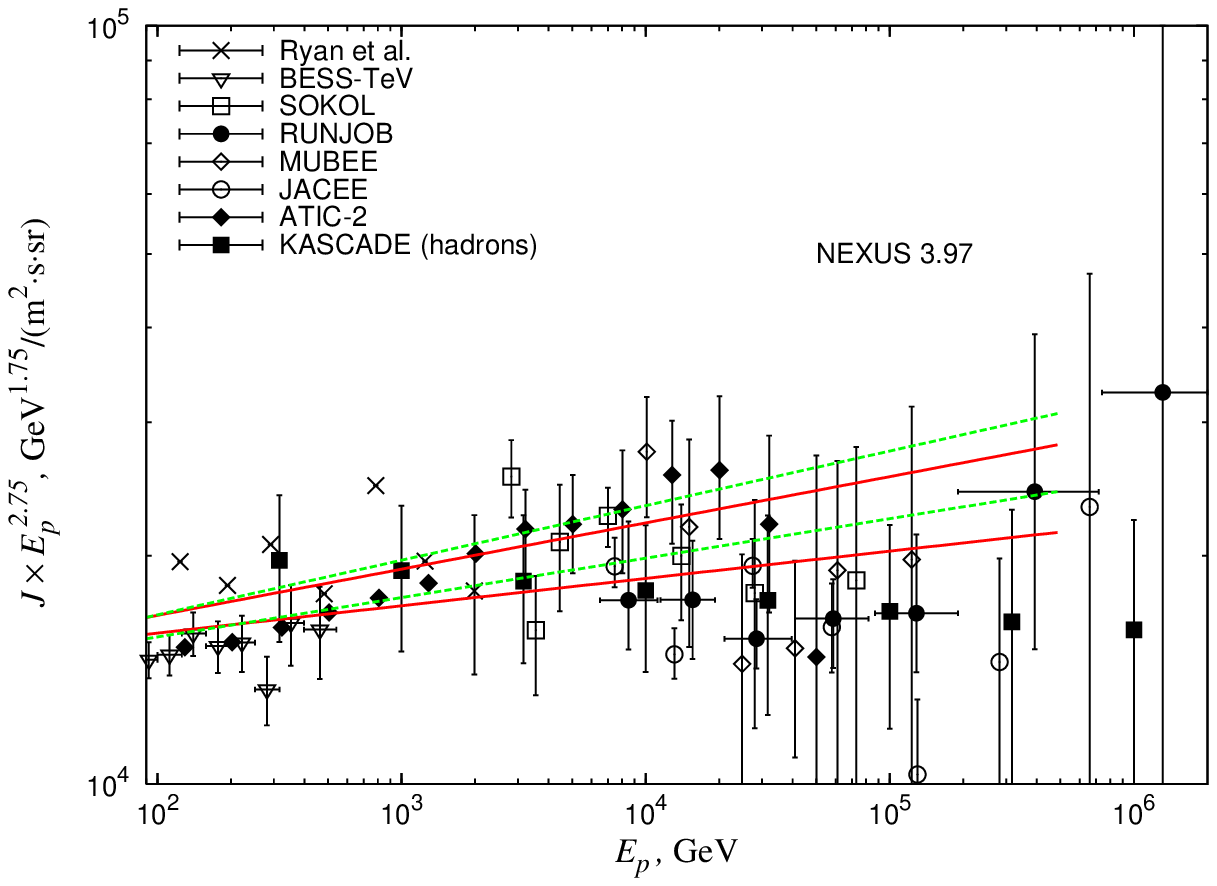}
\includegraphics[width=0.49\textwidth]{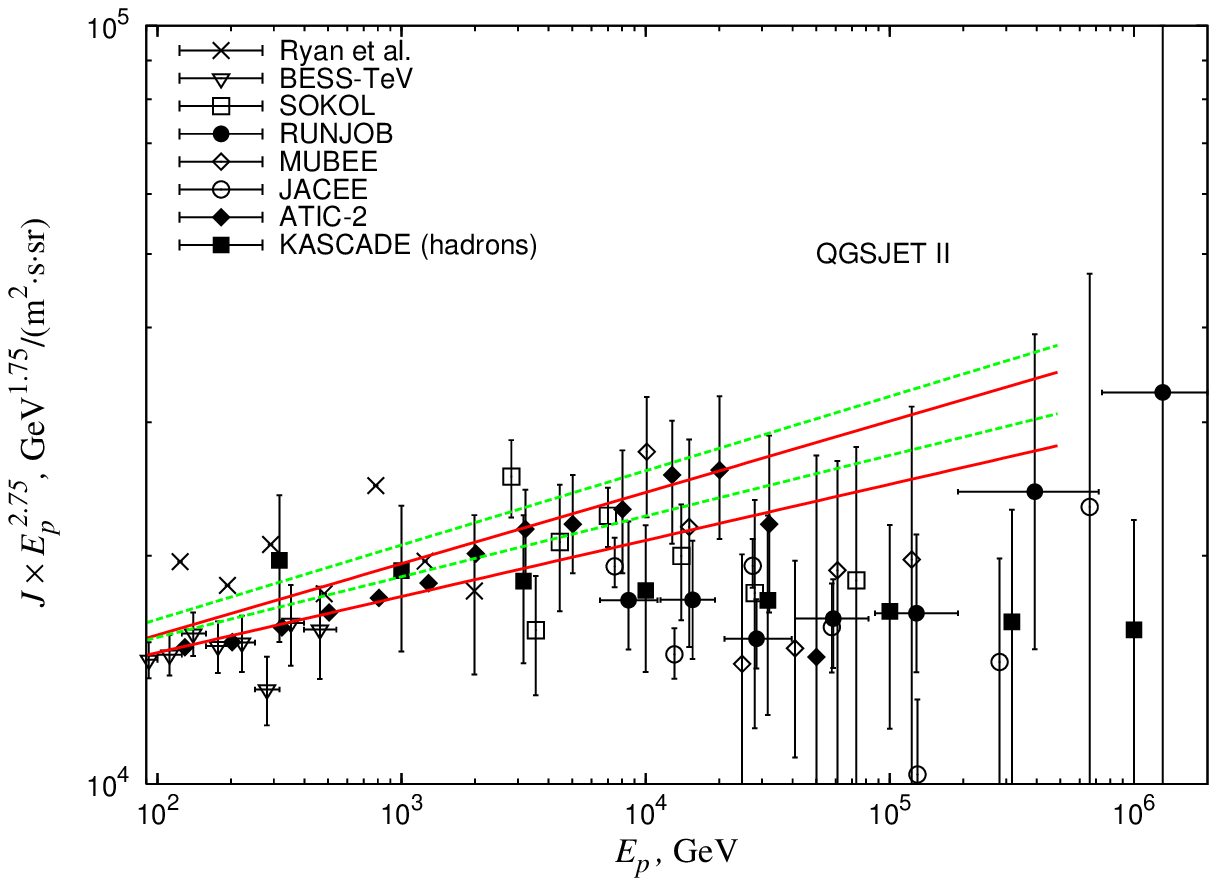}
\caption{(color online). Primary proton spectrum. Experimental data:
\cite{ryan}~Ryan et al., \cite{bess_tev}~BESS-TeV, \cite{sokol}~SOKOL,
\cite{runjob_apj2005}~RUNJOB, \cite{zatsepin}~MUBEE, \cite{jacee}~JACEE,
\cite{atic2}~ATIC-2,
\cite{kascade_p2004}~KASCADE~(hadrons). 
Solid lines and dashed lines show primary proton spectra for muon flux
parametrizations~(\ref{eq:bogdanova}) and~(\ref{eq:mbogdanova}) correspondingly.
Upper and lower lines reflect uncertainty in helium flux according
to~\cite{gaisser2002}.}

\label{fig:primp}
\end{figure*}

\begin{figure}
\centering\includegraphics[width=0.48\textwidth]{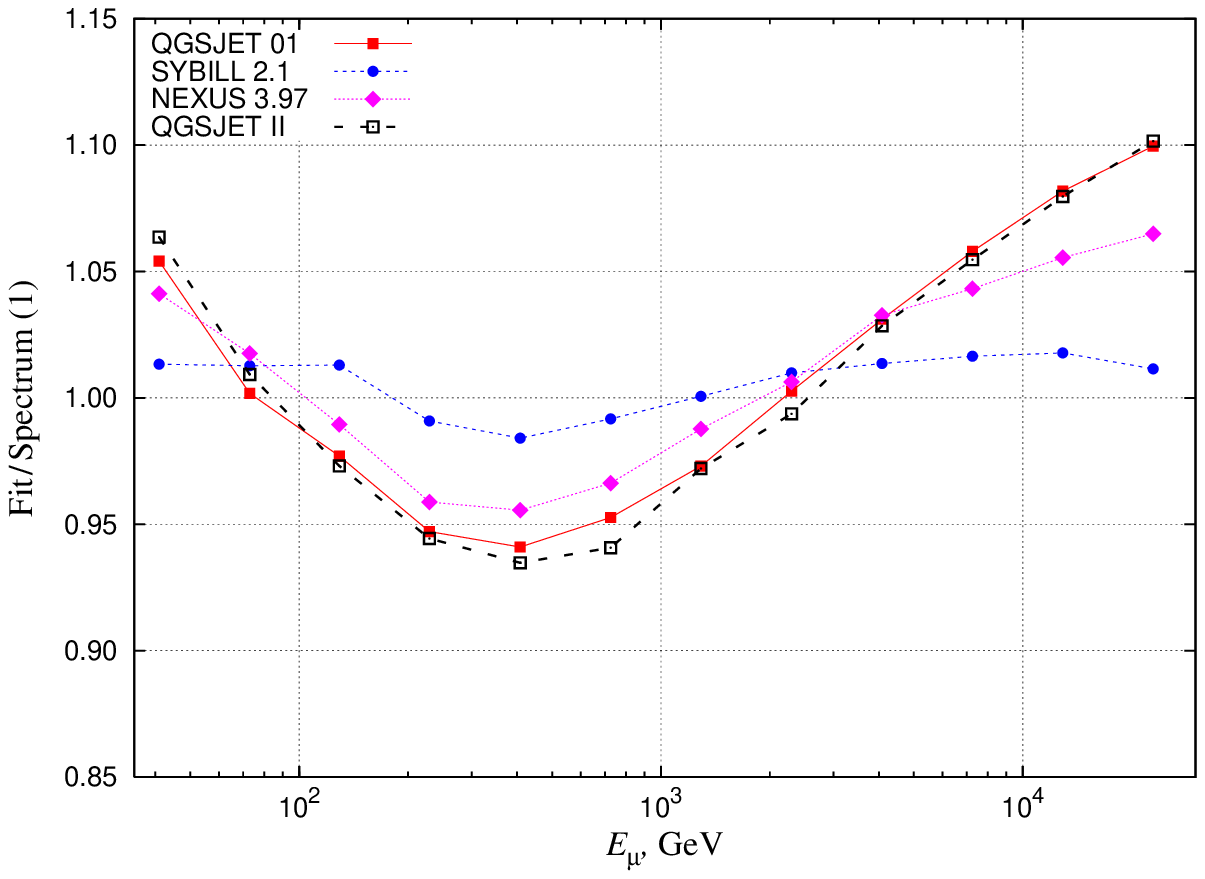}
\centering\includegraphics[width=0.48\textwidth]{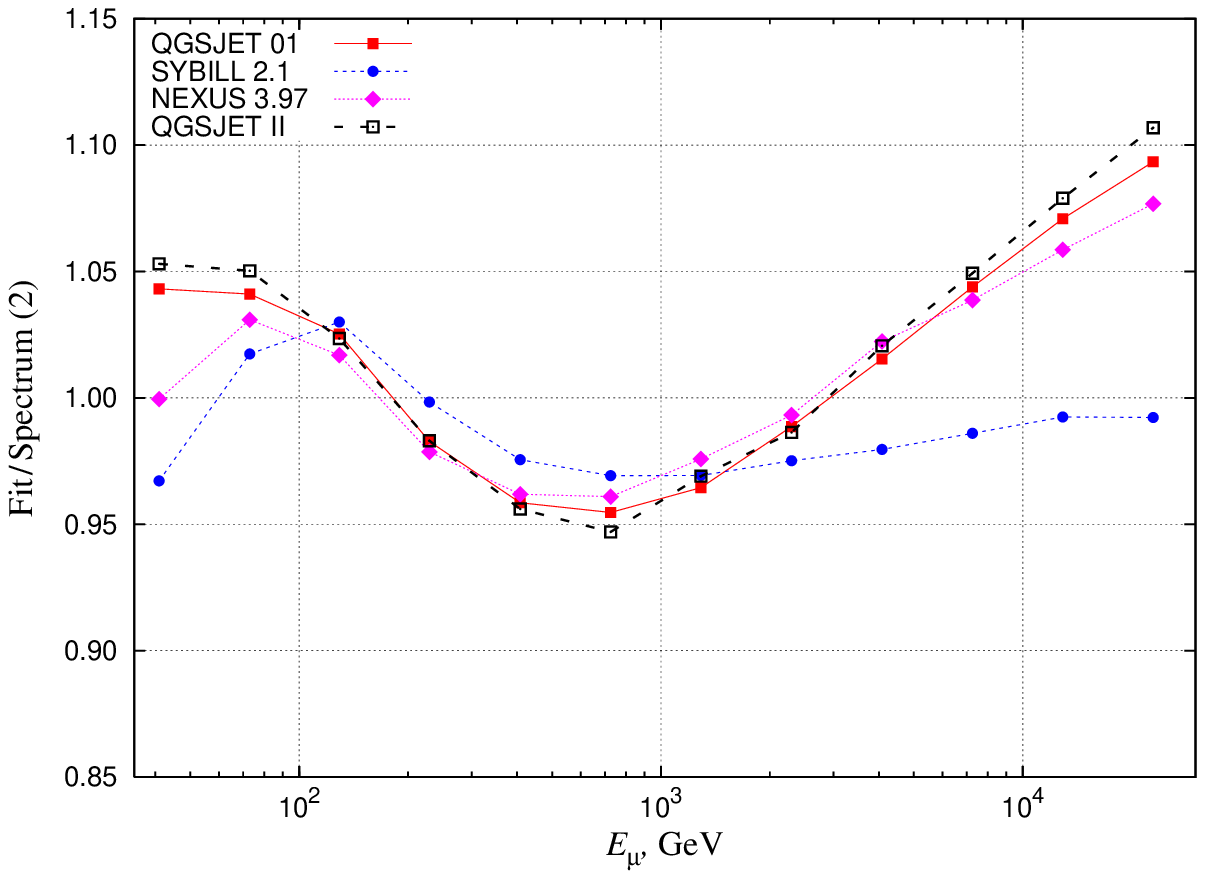}
\caption{(color online). Ratios of muon fluxes, obtained for power-law 
primary proton spectra, given in Fig.~\ref{fig:primp}, to muon flux
parameterizations~(\ref{eq:bogdanova}) and (\ref{eq:mbogdanova}).}
\label{fig:fitratio}
\end{figure}

\begin{figure*}
\centering\includegraphics[width=0.49\textwidth]{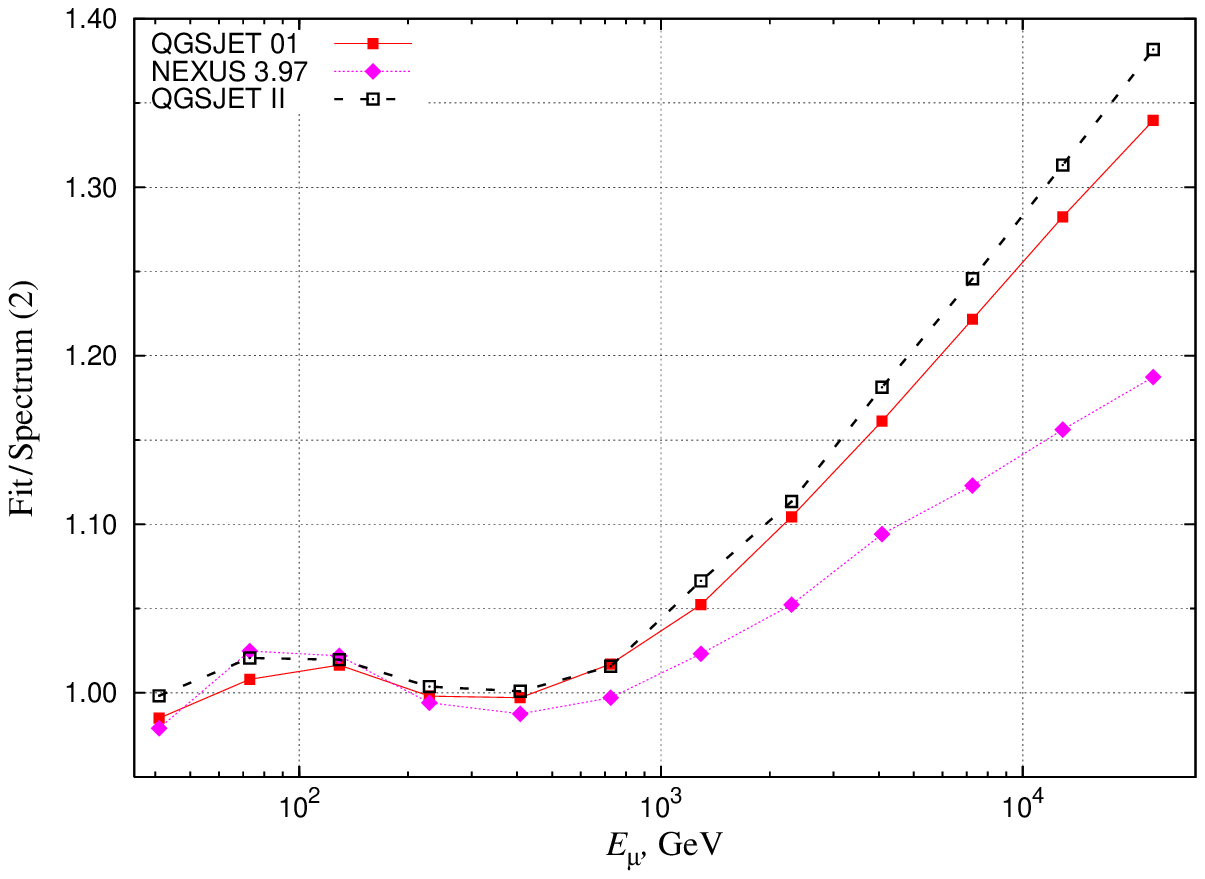}
\includegraphics[width=0.49\textwidth]{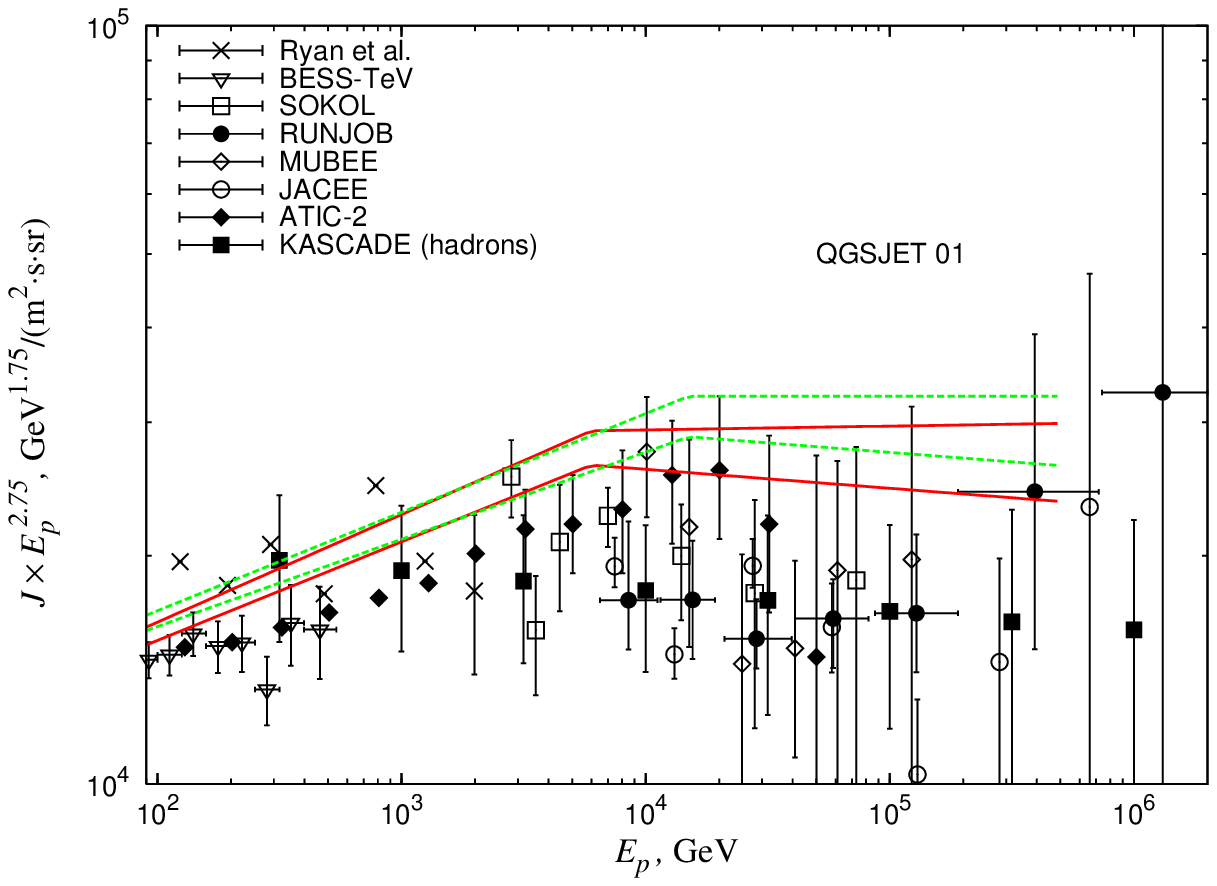}
\includegraphics[width=0.49\textwidth]{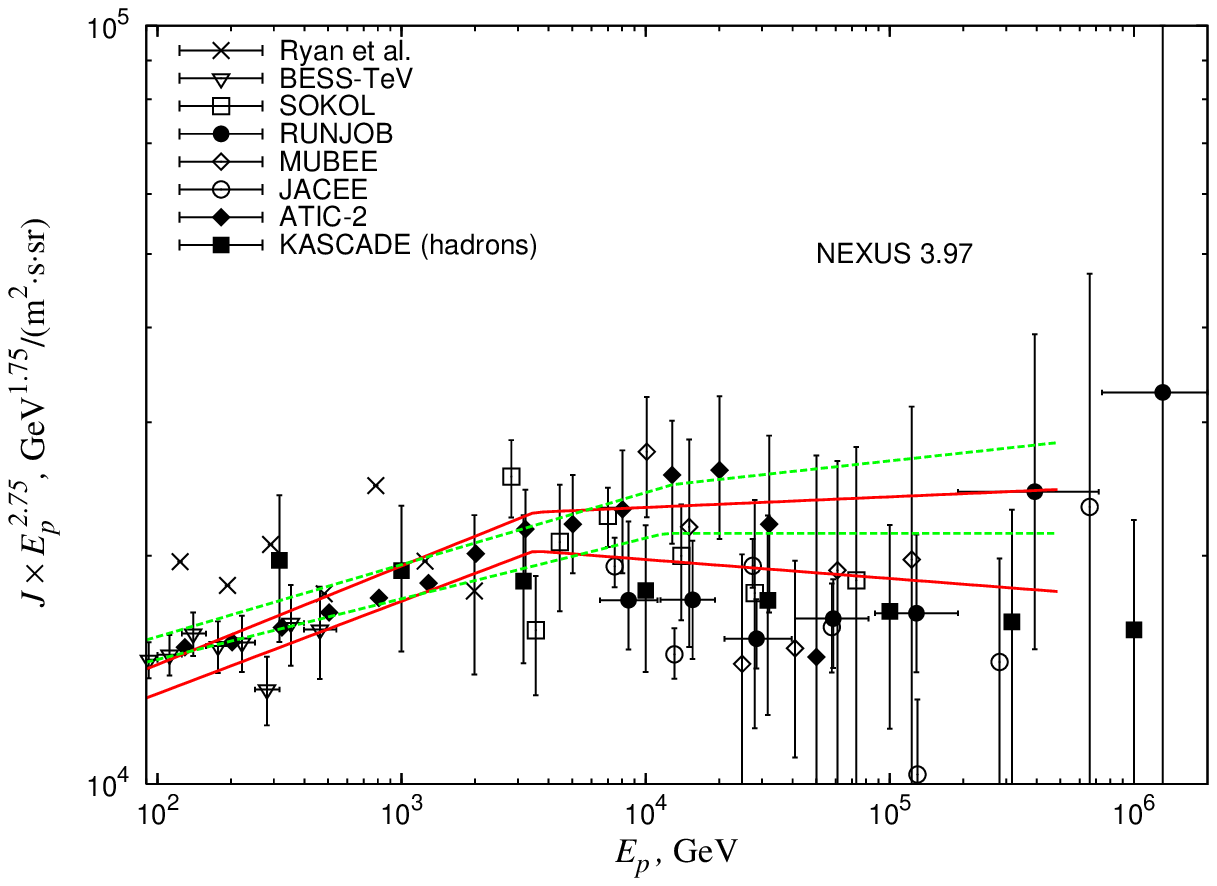}
\includegraphics[width=0.49\textwidth]{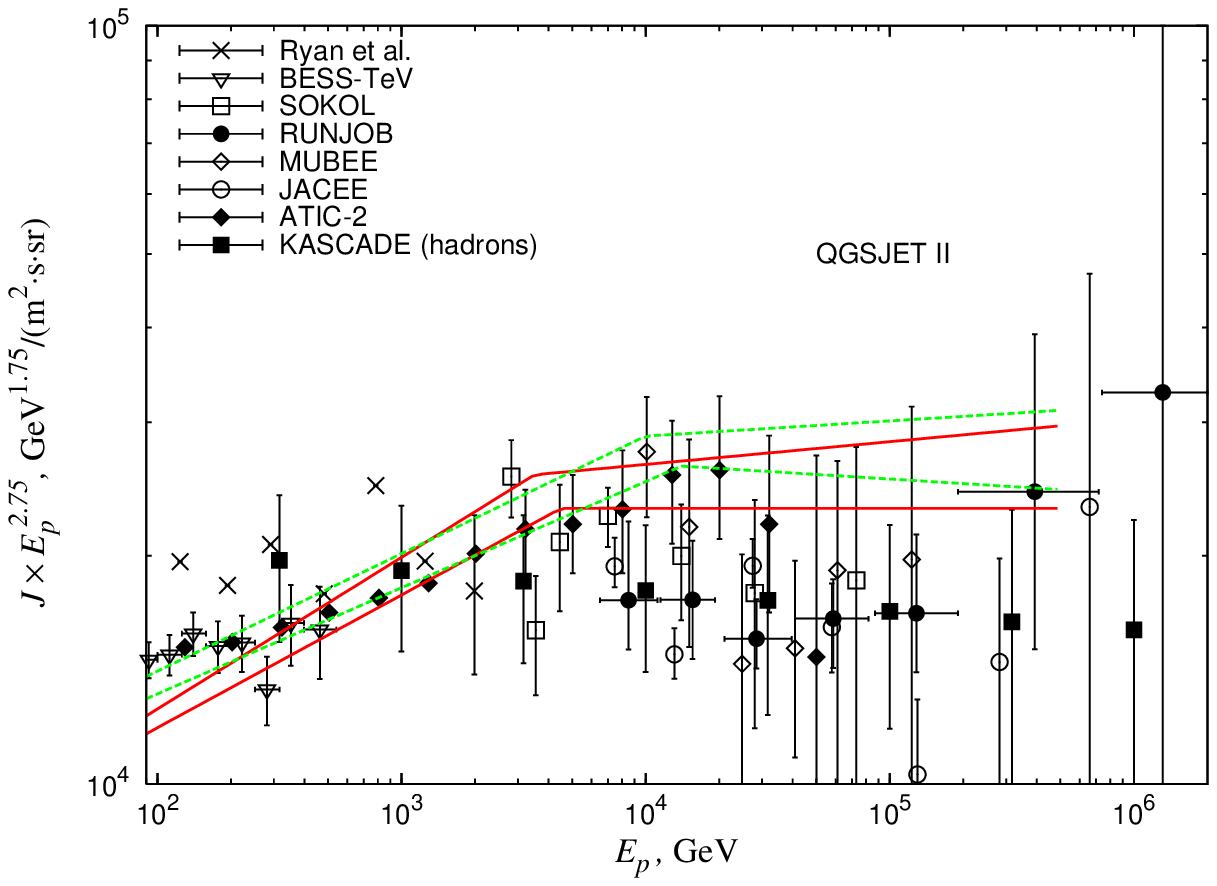}
\caption{(color online). Upper left panel shows ratios of muon fluxes, 
obtained for primary proton spectra with constant power indices equal to those
before break, to muon flux parametrization~(\ref{eq:mbogdanova}). Other
designations as in Fig.~\ref{fig:primp}.}
\label{fig:primpbr}
\end{figure*}

First, let us consider an attempt to minimize maximal deviation  from
spectra~(\ref{eq:bogdanova}),\,(\ref{eq:mbogdanova}) for $E_\mu=40$~GeV--10~TeV
in assumption, that primary proton spectrum can be described by single power law
function $J_p=AE^{-\gamma_p}$ in the entire energy range 100~GeV--500~TeV. In
Fig.~\ref{fig:primp} and in Table~\ref{tab:pp} one can find the results of this
reconstruction, upper and lower lines for each muon spectrum parametrizations in
the Fig.~\ref{fig:primp} correspond to low and high helium flux
fits~\cite{gaisser2002}. As expected, the obtained spectra are flatter, than
measured by RUNJOB and JACEE groups, but, except for QGSJET~01 model, agree well
with ATIC-2 results~\cite{atic2_RCRC2007}. Figure~\ref{fig:fitratio} shows, that
in this case it is possible to achieve agreement with the fitted muon spectrum
within 10\%, but only SIBYLL~2.1 reproduces its shape correctly, with other
models it is not possible to get right muon spectrum slope variation. We should
also note, that this behavior turned out to be insensitive to the choice of
helium flux parametrization. As a result, there is a dip in the ratio of the muon
flux, obtained from fitting of proton spectrum, to the
parameterizations~(\ref{eq:bogdanova}),\,(\ref{eq:mbogdanova}) for energies
around 1~TeV, where the underground data were already underestimated
(overestimation of muon flux at higher energies does not compensate this effect
completely), and growth in small energies range, which brings to contradiction
with low energy $E_\mu<40$~GeV data. Parametrization~(\ref{eq:mbogdanova}) was
also projected to make muon spectrum slope variation less sharp, but this lead to
problems with its fitting in $\sim100$~GeV energy range, as it is seen from 
Fig.~\ref{fig:fitratio} (right panel) for SIBYLL~2.1 and NEXUS~3.97 models. The
problem with muon spectrum shape matching is better illustrated by upper left
panel of Fig.~\ref{fig:primpbr}, where it is shown, that correct reproduction of
muon spectrum for energies below 1~TeV with single power law proton spectrum
leads to appreciable overestimation of muon flux at higher energies. There are
three possible explanations or solutions of this problem. First, the discrepancy
can be completely removed by choice of appropriate interaction parameters, e.g.
similar to those in SIBYLL~2.1. Another argument, which can be given is that the
data on muon flux for energies above 1~TeV are not so definite to claim their
inconsistency with the calculations, but it does not look well supported by
underground data (see Fig.~\ref{fig:strock} and
calculations~\cite{bugaev,ya_yaf2006_zemlya_eng}). And the last possibility is to
assume, that primary proton spectrum is not monotonous and either has sharp break
or slowly changing exponent $\gamma_p$. Let us consider the latter assumption,
which finds experimental~\cite{atic2_RCRC2007} and 
theoretical~\cite{zatsepin_3cmodel,ya_anomal2005_ijmpa,ya_weihai_anomal}
justifications, in more detail. The results for the simple case with break
(Fig.~\ref{fig:primpbr}), which allows to achieve correct description of muon
spectrum shape with right asymptotic and deviation in flux value $<3$\%,  show,
that small difference between spectra~(\ref{eq:bogdanova})
and~(\ref{eq:mbogdanova}) results not only in different proton intensities, but
also in break positions. The latter lies for
parametrization~(\ref{eq:mbogdanova}) in the primary energy range 10--15~TeV, the
change in power index reaches appreciable values up to
$\Delta\gamma_p\approx0.15$ for QGSJET~01 and QGSJET~II models
(Table~\ref{tab:pp}). Proton spectra, obtained from muon
flux~(\ref{eq:mbogdanova}), with QGSJET~II and NEXUS~3.97 models are in the best
agreement with ATIC-2 data, while SIBYLL~2.1 provides intermediate between ATIC-2
and emulsion chambers experiments slope value. Spectra, reconstructed from
parametrization~(\ref{eq:bogdanova}), have breaks at 3--6 TeV and in case of
QGSJET~II proton flux poorly agrees with experiments at primary energies around
100~GeV. Evidently, the latter problems are explained by too low, in comparison
with underground data, muon flux and this parametrization is considered here
mostly for estimation of sensitivity of primary spectrum features to the choice
of reference muon flux. 

It is necessary to note, that due to low sensitivity of differential
muon flux to helium and heavier groups of primary nuclei it is
impossible to derive any conclusions on presence of the break in these
PCR components. For illustration let us consider example of
calculations for QGSJET~01 and high helium flux, where the break in
proton spectrum is positioned at $E_\mathrm{br}=15$~TeV and change of
power index is equal to 0.14 (see
Table~\ref{tab:pp}). Introduction of rigidity-dependent break in He spectrum at
$2E_\mathrm{br}$ per nucleus of the same value $\Delta\gamma=0.14$
gives remarkable discrepancy between calculated muon spectrum and
parameterization~(\ref{eq:mbogdanova}) only for energies above 7~TeV, which
reaches 10\% at 20~TeV. To correct this asymptotic behavior it
suffices to reduce $\Delta\gamma$ to 0.11 simultaneously for protons
and helium without change of the break position, and thus we get
proton spectrum lying well within corridor between parameterizations
for high and low helium fits, shown in Fig.~\ref{fig:primpbr}. Hence,
this corridor covers all possible cases of He flux behavior (with or
without break), provided the helium flux stays within limits, given
in~\cite{gaisser2002}.

Summarizing we can say, that primary proton spectrum shape turns out to be
sensitive to the choice of interaction model and allows presence of break at
10--15~TeV with $\Delta\gamma_p$ up to 0.15, which can be slightly softened
though, if to allow presence of the same break in other PCR components spectra.

\begin{table}
\begin{center}
\captionsetup{labelsep=period}
\caption{Parameters of primary proton spectrum 
fits $J_p=AE^{-\gamma},
(\text{cm}^2\cdot\text{s}\cdot\text{sr}\cdot\text{GeV})^{-1}$, for muon spectrum
parametrization~(\ref{eq:mbogdanova}).}
\label{tab:pp}
\begin{ruledtabular}
\begin{tabular}{lccccccc}

\multicolumn{8}{c}{Low helium flux}\\
\hline
      & \multicolumn{2}{c}{No break} & \multicolumn{5}{c}{With break} \\
\cline{4-8}
Model & \multicolumn{2}{c}{} & \multicolumn{2}{c}{Below break} &Break& \multicolumn{2}{c}{Above break} \\
%\cline{2-3}\cline{4-5}\cline{7-8}
          & $A$ &$\gamma$ & $A$ &$\gamma$ & (TeV)  & $A$ &$\gamma$ \\
\hline
QGSJET 01 & 1.42 & 2.675 & 0.93 & 2.620 & 15 & 3.25 & 2.750\\
QGSJET II & 1.04 & 2.650 & 0.69 & 2.595 & 10 & 2.39 & 2.730\\
NEXUS 3.97& 1.22 & 2.680 & 1.01 & 2.655 & 13 & 1.78 & 2.715\\
SIBYLL 2.1& 1.29 & 2.695 &      &       &      &      &      \\
\hline
\hline
\multicolumn{8}{c}{High helium flux}\\
\hline
      & \multicolumn{2}{c}{No break} & \multicolumn{5}{c}{With break} \\
\cline{4-8}
Model & \multicolumn{2}{c}{} & \multicolumn{2}{c}{Below break} &Break& \multicolumn{2}{c}{Above break} \\
%\cline{2-3}\cline{4-5}\cline{7-8}
          & $A$ &$\gamma$ & $A$ &$\gamma$ & (TeV)  & $A$ &$\gamma$ \\
\hline
QGSJET 01 & 1.45 & 2.690 & 0.95 & 2.635 & 15 & 3.65 & 2.775\\
QGSJET II & 1.08 & 2.670 & 0.69 & 2.610 & 14 & 3.18 & 2.770\\
NEXUS 3.97& 1.25 & 2.700 & 1.01 & 2.670 & 12 & 2.14 & 2.750\\
SIBYLL 2.1& 1.37 & 2.720 &      &       &      &      &      \\

\end{tabular}
\end{ruledtabular}
\end{center}
\end{table}

\section{Hadrons}

\begin{figure*}
\centering\includegraphics[width=0.49\textwidth]{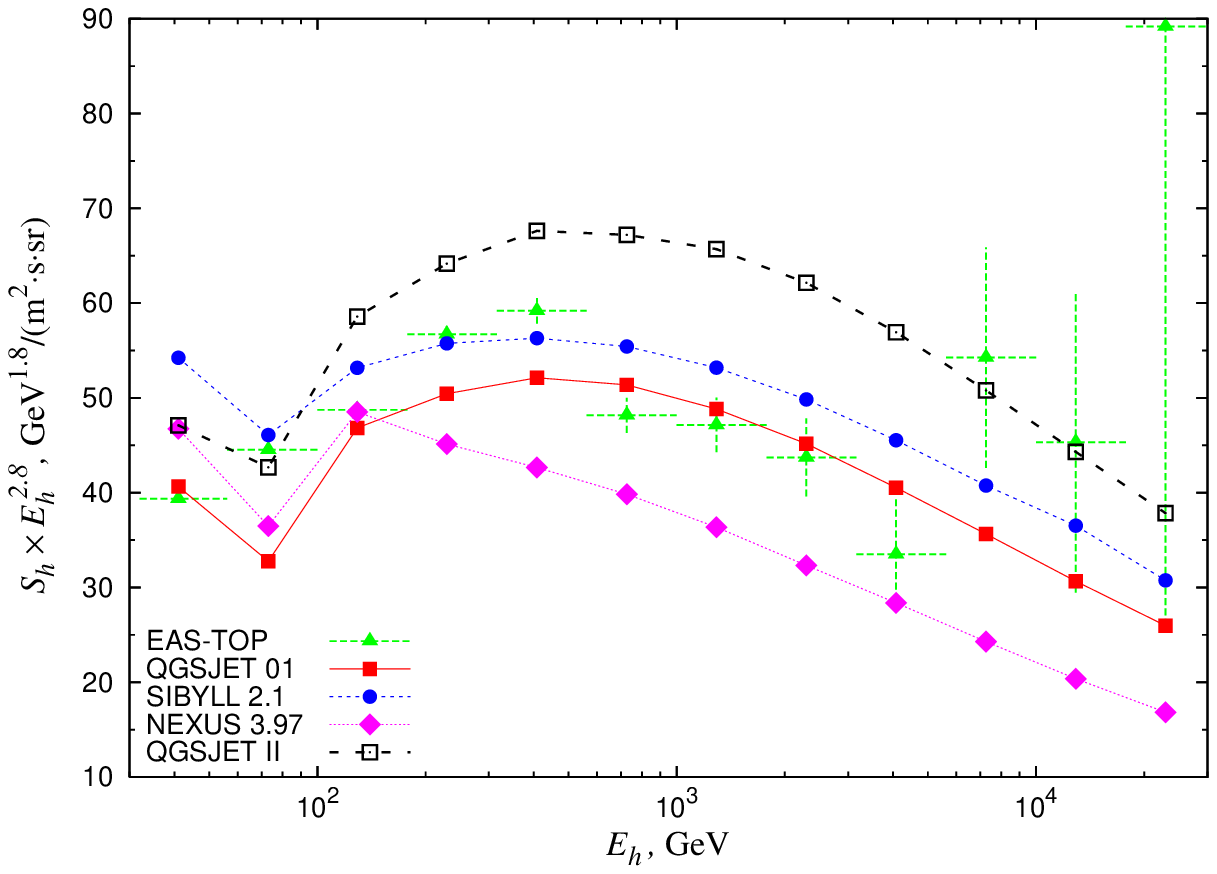}
\centering\includegraphics[width=0.49\textwidth]{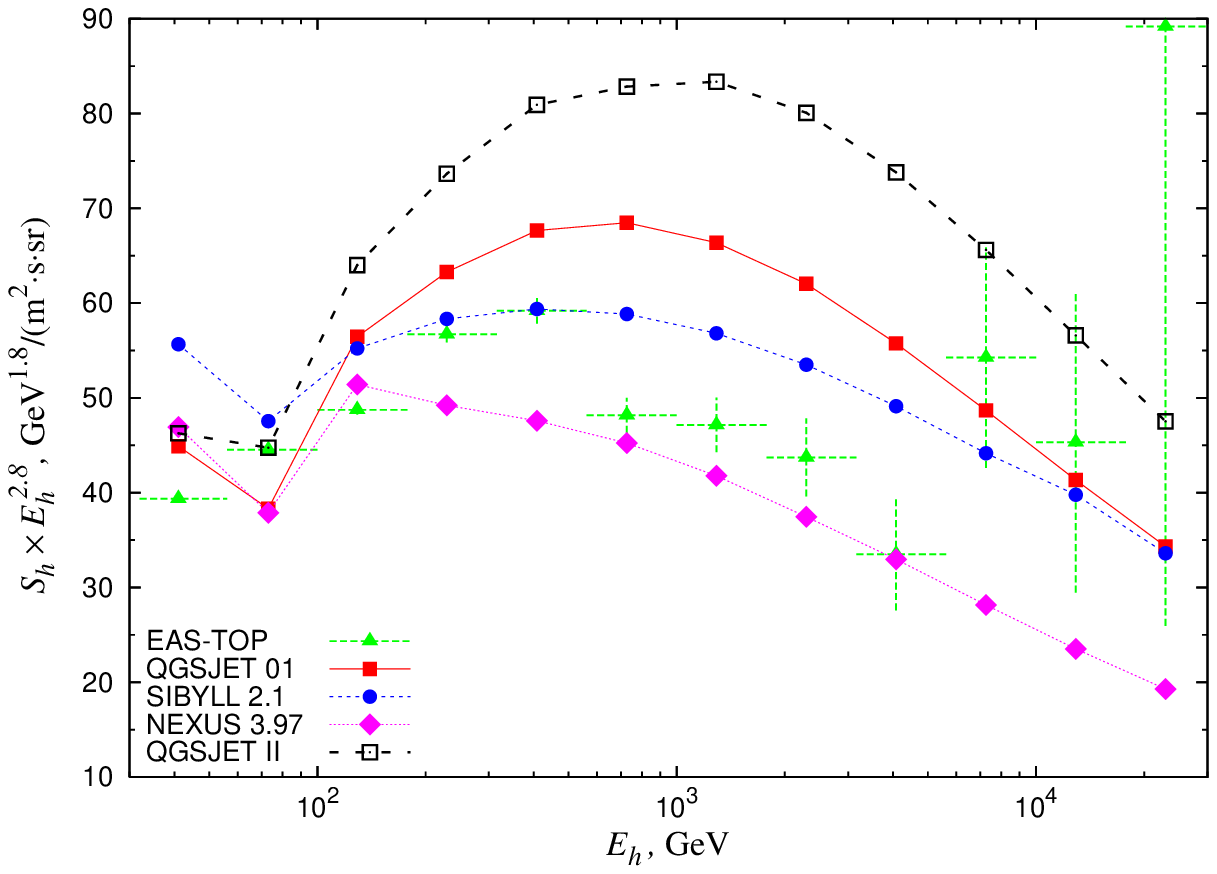}
\caption{(color online). Hadron spectra at the EAS-TOP depth $t=820$~g/cm${}^2$. 
{\em Left:} calculations for primary spectra from~\cite{gaisser2002} with high
helium flux. {\em Right:} calculations for proton spectra from
Fig.~\ref{fig:primpbr} (for SIBYLL~2.1 from Fig.~\ref{fig:primp}).}
\label{fig:eastop}
\end{figure*}

\begin{figure*}
\includegraphics[width=0.49\textwidth]{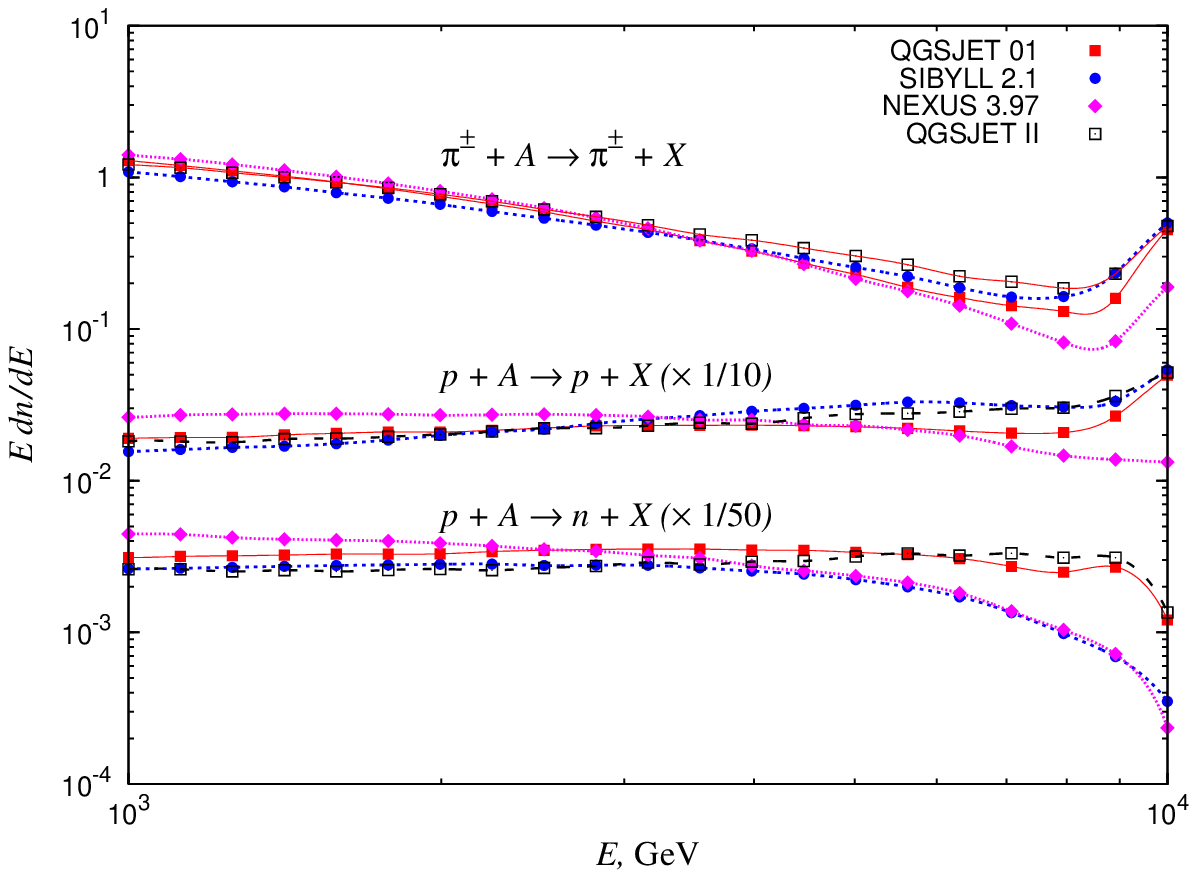}
\includegraphics[width=0.49\textwidth]{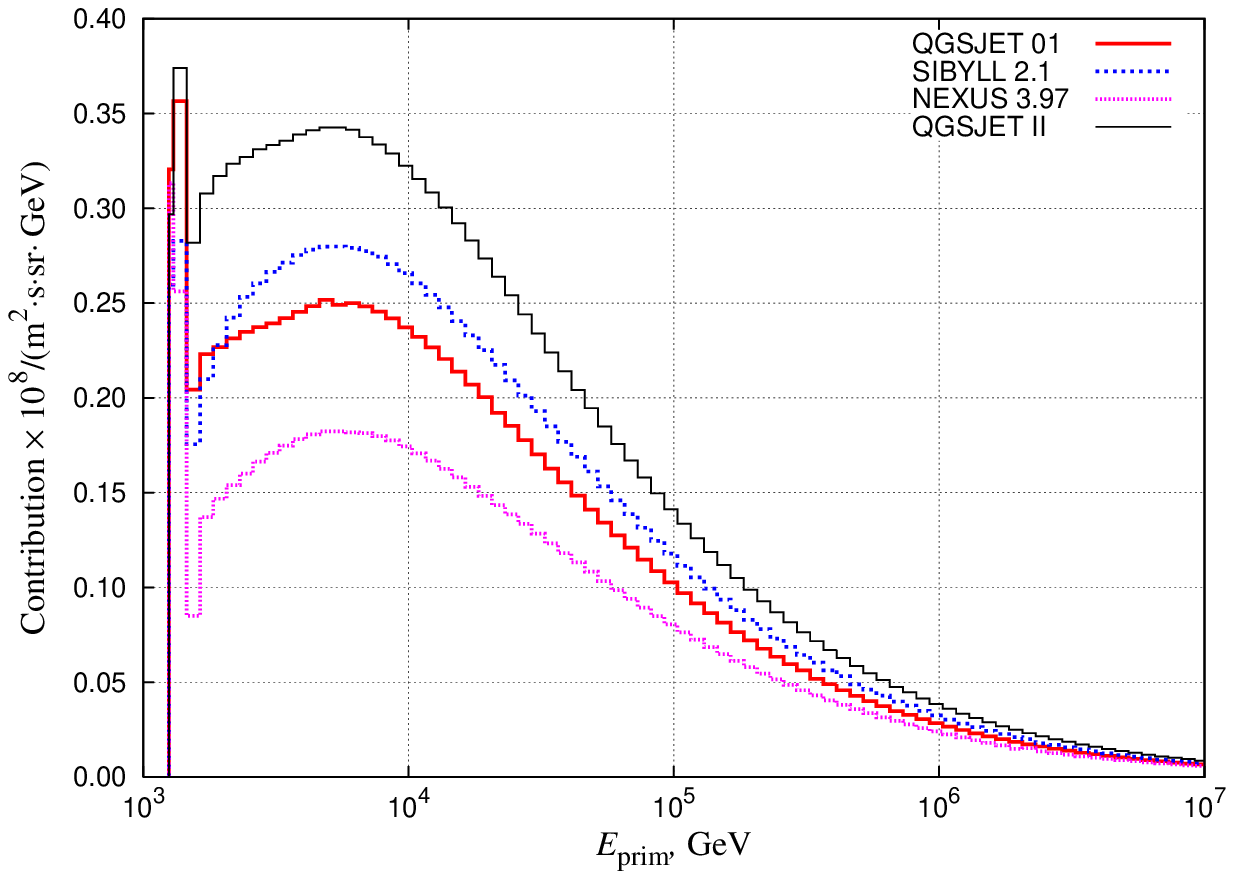}
\caption{(color online). {\em Left:} inclusive spectra $\pi^\pm+A\to \pi^\pm+X$, $p+A\to p+X$ 
(scaled down by 10), $p+A\to n+X$ (scaled down by 50) for incident particles with
energy 10~TeV. {\em Right:} contribution of primary particles with energies
$E_\text{prim}$ to the hadron differential spectrum $S_h(E_h)$ at 820~g/cm$^2$
for $E_h=1.29$~TeV.}
\label{fig:hinclus}
\end{figure*}

\begin{figure*}
\includegraphics[width=0.49\textwidth]{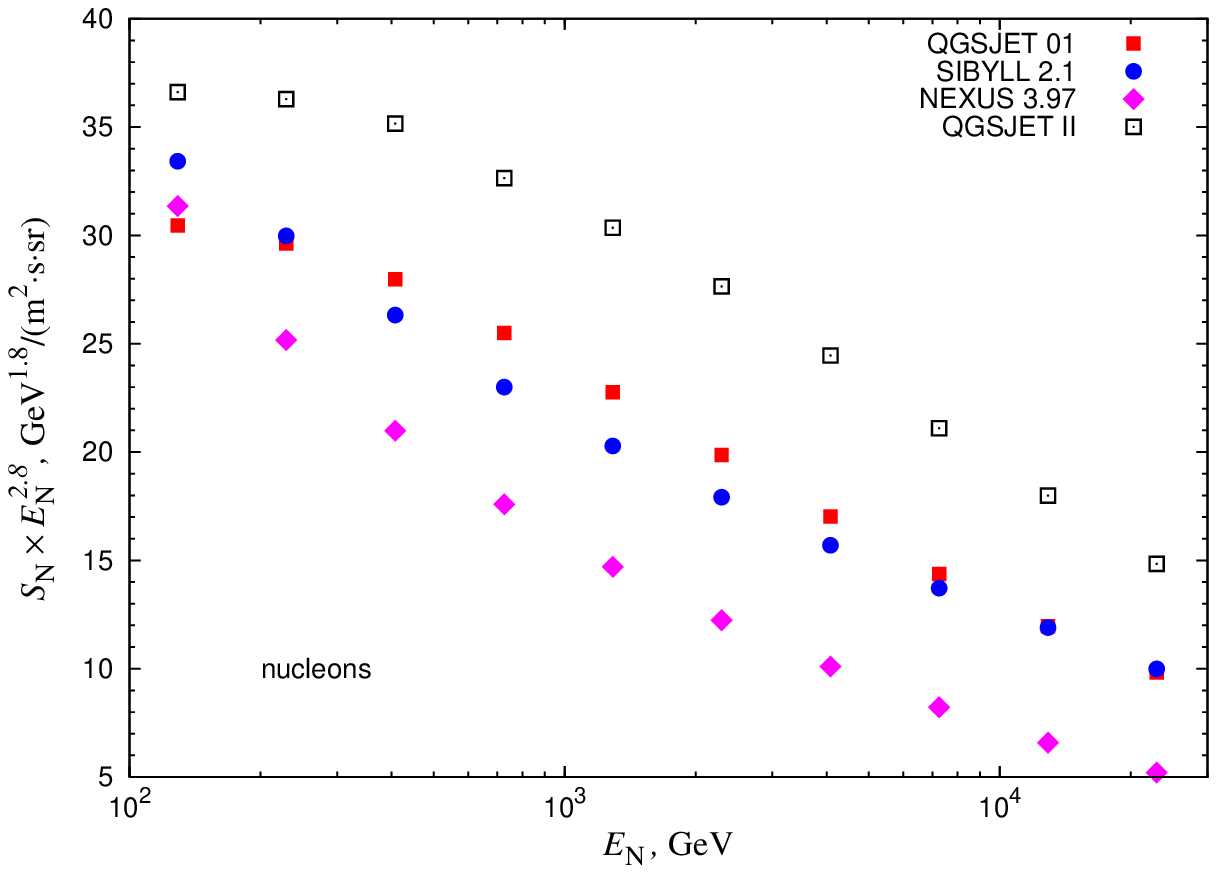}
\includegraphics[width=0.49\textwidth]{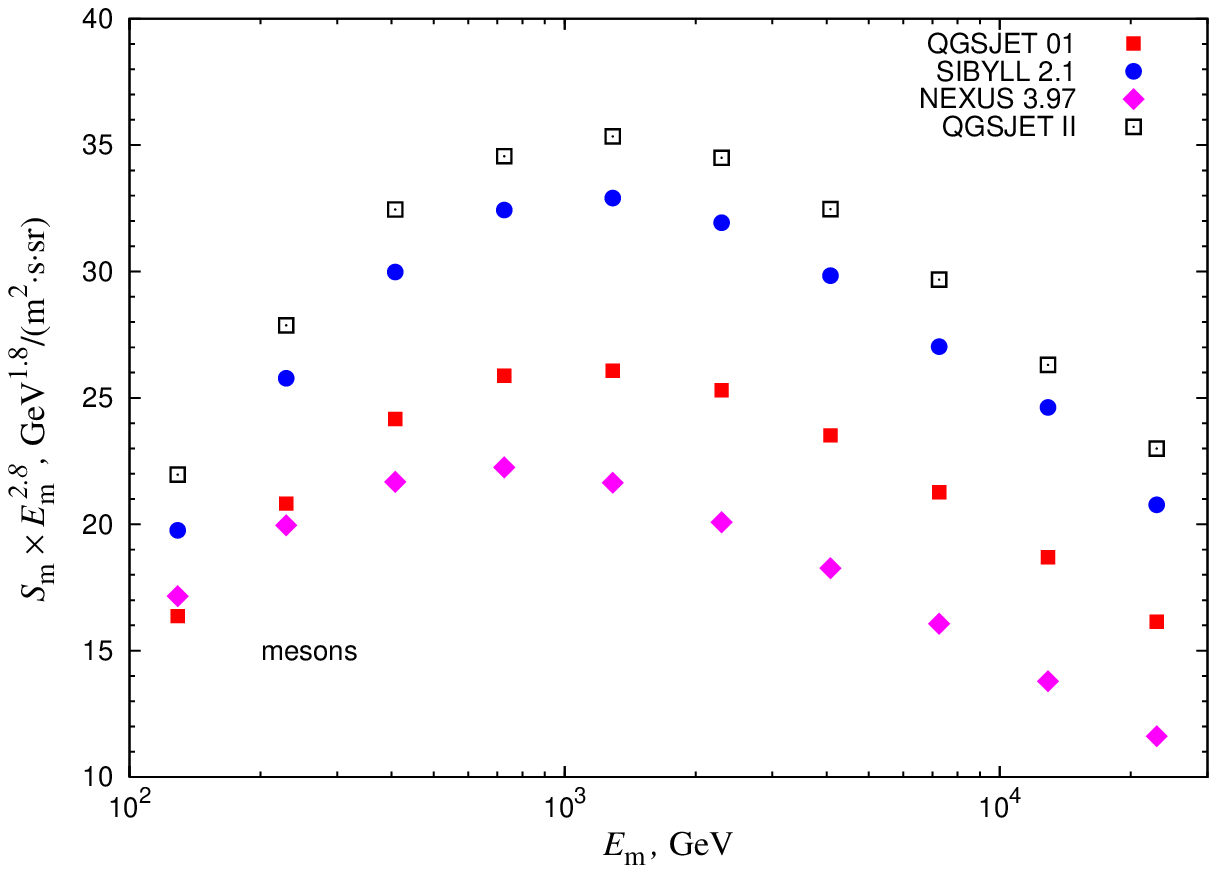}
\caption{(color online). Spectra of nucleons (left) and mesons (right) at 820 g/cm${}^2$ 
for primary spectra~\cite{gaisser2002} with high helium flux.}
\label{fig:mesnuc_sp}
\end{figure*}

Comparison of our calculations of hadron spectrum for the primary spectra
from~\cite{gaisser2002} (high helium flux) with the recent measurements,
performed by EAS-TOP collaboration~\cite{eas_top_p}, is presented in 
Fig.~\ref{fig:eastop} (left panel). First, let us note the following facts.
Below 100~GeV all calculated spectra have breaks, caused by non-perfect
matching of low-energy interaction model GHEISHA to the high-energy models.
Shape of the measured hadron spectra also breaks at energies above 4~TeV and
the data become less definite, thus in the forthcoming analysis we are going to
consider the data for energies from 129~GeV to 4~TeV. For these energies
QGSJET~01, QGSJET~II and SIBYLL~2.1 quite reasonably reproduce the shape of the
measured hadron spectrum, NEXUS~3.97 leads to spectrum with almost constant
power index. One can see, that the most consistent description of the data for
specified energies provide QGSJET~01 and SIBYLL~2.1. In contrast with the muons
there are no energy range, where the models agree on the hadron fluxes and the
reasons of this disagreement are not as simply to point out as in the case with
muons. The most important characteristics in this analysis are total inelastic
cross section, determining chances of primary particle to survive, shapes of
inclusive spectra $p+A\to p+X$, $p+A\to n+X$, $\pi^\pm+A\to \pi^\pm+X$ in the
very forward region, responsible for substantial process of leading particles
production (see Fig.~\ref{fig:hinclus} for the listed spectra). Let us briefly
outline the major conclusions, which one may come to in the given situation.
NEXUS~3.97 gives the lowest fluxes as of hadrons in total, so of nucleons and
mesons (Fig.~\ref{fig:mesnuc_sp}), and this happens in spite of the lowest
inelastic cross-section values. Inclusive spectrum $p+A\to  p+X$ immediately
helps to figure out, that incident protons in NEXUS~3.97 have  comparably low
chances to save most of their energy in collision and this leads to such low
nucleon flux, the same can be said about meson flux and production  of pions by
pions. Similarly, from comparison of the inclusive spectra, it can  be easily
understood, why QGSJET~II gives the highest hadron flux. Note, that  SIBYLL~2.1
concedes to QGSJET~II in hadron intensity mostly because of less  effective
production of leading neutrons in $p$-air collisions and due to the  larger
total interaction cross-section.

Thus, from analysis of the data on hadron flux it is difficult to imply any
strict constraints on inclusive spectra shapes, since mechanism of hadron
spectrum formation is more sophisticated than that in the case of muon
spectrum. SIBYLL~2.1 and QGSJET~01 display quite different behaviors of the
relevant inclusive spectra and total interaction cross-sections, but both
models almost equally succeed in description of the EAS-TOP data (i.e. produce
close hadron fluxes).

The given standard approach to analysis of situation, in fact, is of little
sense, since it is based on assumption about validity of primary spectra in form
of fits from~\cite{gaisser2002}, which was called into question in our previous
discussion. In this case it is logical to analyze interaction models
self-consistency, i.e. their ability to give correct estimates of several CR
components at once. Provided we know behavior of primary proton spectra for every
model, required to match the data on muon flux, we may check how these proton
spectra agree with the data on hadrons. In Fig.~\ref{fig:eastop} (right panel) we
give hadron intensities, calculated for primary proton spectra with breaks from
Fig.~\ref{fig:primpbr} (for SIBYLL~2.1 see Fig.~\ref{fig:primp}), corresponding
to muon spectrum parametrization~(\ref{eq:mbogdanova}). After increase of primary
nucleon flux, dictated by the data on muons, one can see, that the best agreement
with EAS-TOP measurement provide NEXUS~3.97 and SIBYLL~2.1. It would be
interesting to note that two models with different philosophies and inclusive
spectra give the most self-consistent results on muons and hadrons, but, of
course, this conclusion must be taken with much care, since it is based on the
single set of data and we have only indirect indications on the accuracy of this
set, e.g. such as agreement of primary proton fluxes, obtained by EAS-TOP and
KASCADE teams (the latter is derived from flux of unaccompanied
hadrons~\cite{kascade_p2004}). If to try to perform the same analysis with the
variety of the data, obtained at different atmospheric altitudes and zenith
angles, no consistent notions of such kind will be obtained as can be easily seen
from calculations, presented in~\cite{kochanov_2008}.

\section{Conclusions}
The progress in CR and high-energy physics, achieved during last 10--15 years
allowed to turn from statements about satisfactory (qualitative) concordance
between different kinds of data to investigation of more fine effects. As an
example, we managed to show that reconstructed from the data on vertical muon
flux primary proton spectra have not only expected interaction model dependent
intensities, but also model-dependent shapes. It is demonstrated, that
application of QGSJET~01, QGSJET~II and NEXUS~3.97 models brings to proton
spectrum with break at 10--15~TeV and power index $\gamma_p$ before break close
to that, measured in ATIC-2 experiment. Nevertheless one can see, that absolute
proton flux for QGSJET~01 is hardly compatible with any data of direct
experiments, and the break for all these three models is more moderate, compared
to what can be inferred from ATIC-2 data, which though become less definite
right in the break region. On the other hand SIBYLL~2.1 allows to reproduce shape of
the muon spectrum with single power law proton spectrum, which is in reasonable
agreement with both emulsion chamber and ATIC-2 data within experimental errors. 
Further improvement of our understanding of the situation, which is of primary
astrophysical interest, can be achieved via experimental study as of muon CR
component characteristics with water and ice neutrino telescopes and so of
inclusive spectra $p+A\to \pi^\pm, K^\pm$ in fragmentation region. Reduction of
uncertainties in the latter component with the use of the data on primary
spectra, hadron and muon components, does not look possible, because of 1) poor
correlation between muon and hadron production mechanisms, 2) ambiguity of
existing CR experimental data and 3) possibility to realize self-consistent
description of the data on muons and hadrons with the models, having remarkably
differing inclusive spectra and underlying philosophies.

\section*{ACKNOWLEDGMENTS}
Authors greatly acknowledge CONEX team and personally Tanguy Pierog for their
kind permission to use CONEX cross-section tables and for technical support.
We are grateful to anonymous referee for constructive remarks, which helped us
to improve the manuscript.\\ This work was supported in part by RFBR grant
07-02-01154-a.

\iftrue

\fi

\end{document}